\documentclass[epsfig,12pt, onecolumn]{article}
\usepackage{epsfig}
\usepackage{amsfonts}
\usepackage{axodraw}
\newcommand{\newc}{\newcommand}
\newc{\gsim}{\lower.7ex\hbox{$\;\stackrel{\textstyle>}{\sim}\;$}}
\newc{\lsim}{\lower.7ex\hbox{$\;\stackrel{\textstyle<}{\sim}\;$}}
\newc{\gev}{\,{\rm GeV}}
\newc{\mev}{\,{\rm MeV}}
\newc{\ev}{\,{\rm eV}}
\newc{\kev}{\,{\rm keV}}
\newc{\tev}{\,{\rm TeV}}
\def\ln{\mathop{\rm ln}}

\newc{\mz}{M_Z}
\newc{\mpl}{M_*}
\newc{\mw}{m_{\rm weak}}
\newc{\nr}[1]{N^c_R{}_{#1}}
\usepackage{amsmath}
\def\beq{\begin{equation}}
\def\eeq{\end{equation}}
\def\bea{\begin{eqnarray}}
\def\eea{\end{eqnarray}}
\def\bitem{\begin{itemize}}
\def\eitem{\end{itemize}}
\newc{\ie}{{\it i.e.}}          \newc{\etal}{{\it et al.}}
\newc{\eg}{{\it e.g.}}          \newc{\etc}{{\it etc.}}
\newc{\cf}{{\it c.f.}}
\def\bar#1{\overline{#1}}
\def\vev#1{\left\langle #1 \right\rangle}

\def\abs#1{\left| #1\right|}

\def\inv{^{\raise.15ex\hbox{${\scriptscriptstyle -}$}\kern-.05em 1}}
\def\lbar{{\lower.35ex\hbox{$\mathchar'26$}\mkern-10mu\lambda}} 

\def\eq#1{Eq.(\ref{#1})}

\let\<=\langle
\let\>=\rangle

\let\+=\uparrow

\let\be=\beta
\let\ga=\gamma

\let\ep=\epsilon

\let\la=\lambda
\let\La=\Lambda

\let\si=\sigma

\def\singlespaced{\baselineskip=\normalbaselineskip} 
\begin{document}
\thispagestyle{empty}
\vspace*{.5cm}
\noindent
\hspace*{\fill}{\large OUTP-07-20P}\\
\hspace*{\fill}{\large FERMILAB-PUB-08-014-A}\\
\vspace*{2.0cm}

\begin{center}
{\Large\bf Heavy Dark Matter Through the Higgs Portal}
\\[2.5cm]
{\large John March-Russell$^1$, Stephen M. West$^{1,2}$, Daniel Cumberbatch$^3$ and Dan Hooper$^4$.
}\\[.5cm]

{\it$^1$Theoretical Physics, Department of Physics\\
University of Oxford, 1 Keble Road, Oxford OX1 3NP, UK}
\\[.2cm]
{\it$^2$Magdalen College, Oxford, OX1 4AU, UK}
\\[.2cm]
{\it $^3$Astrophysics Dept., University of Oxford,\\  
Denys Wilkinson Building, Oxford OX1 3RH, UK} 
\\[.2cm] 
{\it $^4$Center for Particle Astrophysics, Fermi National Accelerator Laboratory, Batavia, IL 60510-0500, USA} 
\\[.2cm] 
(January 22, 2008)
\\[1.1cm]

{\bf Abstract}\end{center}
\noindent
Motivated by Higgs Portal and Hidden Valley models, heavy particle dark matter that communicates with the supersymmetric Standard Model via pure
Higgs sector interactions is considered.  We show that a thermal relic abundance consistent with the measured density of dark matter is possible for masses up to $\sim 30\tev$.  
For dark matter masses above $\sim 1\tev$, non-perturbative Sommerfeld corrections to the annihilation rate are large, and have the potential to greatly affect indirect detection signals.    For large dark matter masses, the Higgs-dark-matter-sector couplings are large and we show how such models may be given a UV completion
within the context of so-called ``Fat-Higgs" models.   
Higgs Portal dark matter provides an example of an attractive alternative to conventional MSSM neutralino dark matter that may evade discovery at the LHC, while still being within the reach of current and upcoming indirect detection experiments.

\newpage
\setcounter{page}{1}

\section{Introduction}

Recently, there has been a surge of interest in models where the Standard Model (SM) or the Minimal Supersymmetric Standard Model (MSSM) communicates
with a partially hidden sector via either $Z'$ or Higgs interactions  \cite{HV,W,wells,PW,PW2,HV2}.   These Hidden Valley or
Higgs Portal models provide a stimulating and consistent alternative to the usual model building assumption of a
desert above the weak scale.   Higgs-sector and $Z'$
interactions between the hidden sector and the SM states are special in that they involve
gauge-invariant operators of dimension $d_{O} \leq 4$, and thus can be
induced by physics at arbitrarily high scales with unsuppressed couplings.  In the case of
a $Z'$ the interactions can either occur directly with SM states if they are charged under the $U(1)'$ or,
possibly more interestingly, indirectly due to a kinetic-mixing term, $\ep F_{Y}^{\mu\nu} F'_{\mu\nu}$,
between hypercharge and the new $U(1)$, in which case $\ep$, and thus the
effective size of the SM-hidden sector interaction, can be suppressed \cite{kineticmixing, HV, HV2}.   
On the other hand, in the case of the Higgs-sector interactions of interest to us here, couplings
of the form $|H|^2 s^2$ involving the SM or MSSM Higgs states and new SM gauge singlet
states can be large, especially in the situation where the TeV-scale theory UV-completes not far above the 
weak scale to a strongly interacting theory with light composite states. 

It is interesting to ask whether such models lead to new dark matter candidates with qualitatively
different phenomenology.  In this paper we argue that dark matter communicating with
a supersymmeterized SM purely via Higgs-sector interactions (the Higgs Portal) leads to new
and unusual features.\footnote{Other works which consider aspects of dark matter phenomenology in the context of Hidden Valley or Higgs Portal models are contained in Ref.~\cite{gmsbdm,hiddendm}, while earlier related studies are contained in Ref.~\cite{gaugesingletdm}. }
First, as we will show, the thermal relic abundance in such a scenario can be consistent with the measured density of dark matter for masses as high as $\sim 30\tev$, much larger than are usually considered (while also being consistent with the upper bound on the mass of thermal relic dark
matter derived from unitarity~\cite{GK}).   Second, for dark matter masses above $\sim 1\tev$ non-perturbative Sommerfeld corrections \cite{sommer} to the low-velocity annihilation rate are large.  Several authors have recently recognised the potential importance of these corrections to the dark matter relic density calculations 
\cite{gunion, hisano, profumo, strumia}, which lead to enhanced annihilation rates in the case of attractive interactions.  Even more importantly, as we will argue in detail in a companion paper \cite{companion}, these corrections have the potential to greatly enhance the indirect annihilation signals by factors of up to $10^5$ beyond those predicted without consideration of the Sommerfeld factor, potentially leading to a significant change in the optimal search strategy.   

As well as providing examples in which the dark matter particle is beyond the kinematic reach of the Large Hadron Collider (LHC) but is potentially detectable by indirect and direct dark matter searches, the models presented here are independently motivated by the desire to raise the MSSM upper bound on the lightest Higgs mass, and so relax the current tension with the LEP2 Higgs-mass exclusion limit.  It is also interesting that our models may be given a UV completion in so-called ``Fat-Higgs" models \cite{Harnik:2003rs}\footnote{Other models in a similar class to the Fat Higgs model are discussed in Ref.~\cite{uvcomplete}.}
in which some TeV-scale states are composites of the underlying strong-coupling dynamics.  This UV completion is consistent with both collider constraints and aesthetic requirements such as gauge coupling unification.  This completion is discussed in detail in Section~\ref{fathiggs}.

Furthermore, the existence of partially hidden (secluded) sectors is common in models that attempt to embed the SM within a larger structure.   Well studied examples include higher-rank GUT models, such as those based upon $E_6$ \cite{gut}, and supersymmetry breaking models, in particular the messenger sectors of gauge-mediated supersymmetry breaking models~\cite{gmsbdm}.  More recently, it has been argued that secluded or hidden sectors in the form of Randall-Sundrum-like warped ``throats'' \cite{throats} are a ubiquitous feature of the landscape of string compactifications \cite{ubiquitous}, thus implying that there is not an insignificant probability that a hidden or secluded throat with a mass scale close to the weak scale exists.  In fact, as argued by Patt and Wilczek \cite{PW}, the scales in sectors interacting by Higgs portal interactions are commonly tied together.   

Naturally, if our dark matter candidate is to be the dominant component of the cosmological dark matter, we must ensure that the usual neutralino dark matter candidate of the MSSM leads either to a subdominant relic density or is unstable.  In the case in which $R$-parity is conserved and a neutralino is the lightest supersymmetric particle (LSP), the thermally generated abundance of such a state is in many models well below the measured dark matter density. In particular, wino-like or higgsino-like LSPs annihilate very efficiently, leading to subdominant abundances~\cite{dmreview}. Coannihilations with other supersymmetric states can also deplete the neutralino abundance in many models~\cite{GS}. Alternatively, instead of being a neutralino, the LSP could be a different supersymmetric state, such as a gravitino. Within the context of gauge-mediated supersymmetry breaking, for example, the LSP is typically a light gravitino which constitutes only a very small fraction of the cosmological dark matter abundance. On the other hand, if there exist $R$-parity violating interactions, then the LSP will be unstable thus evading this issue entirely.\footnote{A late-decaying LSP may even be beneficial in that it can correct
the BBN prediction for the $^6$Li to $^7$Li ratio \cite{lithium}.}

Turning to the structure of our paper, in Section~\ref{models} we introduce
our models and explain how they are a modified form of the so-called
Minimal Non-minimal Supersymmetric Standard Model (MNSSM),
while in Section~\ref{som} we give a brief introduction to the physics
of the Sommerfeld enhancement that plays an important role in our calculations.
In Section~\ref{reliccalc} we summarize
the calculation of the relevant dark matter annihilation cross section including the Sommerfeld enhancement and present our results for the relic density.  In Section~\ref{detection}
we briefly discuss the direct and indirect detection of our dark matter candidate, leaving a more detailed study for a companion paper~\cite{companion}.   Section~\ref{fathiggs}, in which we demonstrate
that our models may be given a UV completion in so-called ``Fat-Higgs" models where
the states are composites of underlying strongly coupled dynamics, is somewhat outside
the main development of our paper and may be skipped by readers only interested in dark matter phenomenology.
Finally, our conclusions are given in Section~\ref{conclusions}.

\section{The Supersymmetric Higgs Portal Model}\label{models}

The relevant terms of the model that we wish to study are specified by the superpotential
\beq
W=W_{MSSM}(\mu=0)+\la NH_uH_d+\frac{\la^{\prime}}{2}NS^2+\frac{m_{\tilde{s}}}{2}S^2+... ,
\label{sup1}
\eeq
where $N$ and $S$ are SM singlets and $N$ gets an electroweak-sized scalar vacuum
expectation value (vev).  The term $W_{MSSM}(\mu=0)$ refers to the MSSM
superpotential without the $``\mu"$ term, while the ellipsis denote terms, such as possible tadpoles, that will not be important.  $S$ has an exact non-R
$Z_2$ symmetry which will be unbroken in the vacuum and which leads to a stable relic,
$\tilde{s}$, the fermionic component of the $S$ superfield with mass
$m_{\tilde{s}}$.  Note that $N$ does not have a mass term before electroweak symmetry
is broken.  In fact the masslessness of $N$ before EWSB is not crucial; all that is required
is that the mass of $N$ is small compared to $S$ as we explain in detail below.  We will assume
that the standard neutralino supersymmetric dark matter
candidate is irrelevant, either because $R_p$ is broken, or because its relic density is
subdominant to that of $\tilde{s}$.  This model \eq{sup1} is a simple variation of the model outlined in Refs.~\cite{pil1} and \cite{Ded1}, referred to as the Minimal Non-minimal Supersymmetric Standard Model (MNSSM) in Ref.~\cite{pil1}, where the superpotential has the form $W_{MNSSM}=W_{MSSM}(\mu=0)+\la NH_uH_d+t_2N$, and $t_2$ is a mass dimension two ``tadpole"-term parameter that is in general possible.

The annihilation cross sections determining the number density of our dark matter particle will depend on the couplings, $\la$ and $\la^{\prime}$.  As we will argue in later sections, the most interesting dark matter phenomenology occurs when the coupling $\la^{\prime}$ is large.  Furthermore, for large $\la$ there are important additional contributions to the higgs quartic-self-couplings and the upper bound on the lightest higgs mass is considerably raised. 

For sufficiently large $\la, \la^{\prime}$ couplings the theory \eq{sup1} hits a Landau
pole below the Planck-scale, and so must be considered a low-energy effective theory
with a cutoff $\La$. 
We will argue in Section~\ref{fathiggs} that the above effective theory can result from a limit of the Fat Higgs model \cite{Harnik:2003rs} in which $S$ is a composite meson field of new supersymmetry-preserving strong-interaction dynamics, giving our effective theory a possible and plausible UV completion without tadpole problems, and also with a natural reason to expect large couplings $\la$ and $\la^{\prime}$.  We emphasize
that the Fat Higgs model is only one of many possible examples of a UV completion with large couplings $\la$ and $\la^{\prime}$.  To analyse the dark matter phenomenology it is sufficient to
focus on the effective superpotential in \eq{sup1} rather than that of any particular UV completion.  Although the precise form of the superpotential,
Eq.(\ref{sup1}), arises as a low-energy limit of the Fat Higgs model, other terms, such as a bare
$\mu$-term, a mass term for $N$, or $N^3$ self-interactions can be added to \eq{sup1} without qualitatively changing our results if the final mass of $N$ is parametrically smaller than $m_{{\tilde s}}$ by a factor of  ${\cal O}(10)$.
\footnote{More precisely, the spectrum of neutral $Z_2$-even Higgs scalars arising from $H_u, H_d, N$ after mixing must be such that a state with substantial interaction with the $Z_2$-odd
states $s, {\tilde s}$ has mass which is parametrically small compared to $m_{{\tilde s}}$.}
We assume this in the remainder  of our analysis.

From the superpotential, \eq{sup1}, the Lagrangian terms determining the important interactions and masses in the model are
\beq
\mathcal{L}=\mathcal{L}_{\rm fermion}+\mathcal{L}_{\rm scalar},
\eeq
where,
\bea
\nonumber
\mathcal{L}_{\rm ferm}&=& -\la n\tilde{h}_u \tilde{h}_d-\la \tilde{n}\tilde{h}_uh_d-\la \tilde{n}h_u\tilde{h}_d-\frac{\la^{\prime}}{2}n
\tilde{s}\tilde{s}-\la^{\prime}\tilde{n}\tilde{s}s-\frac{m_{\tilde{s}}}{2}\tilde{s}\tilde{s}+\mbox{h.c.}+...\\
\nonumber
\mathcal{L}_{\rm scal}&=&\abs{\la^{\prime} ns+m_{\tilde{s}}s}^2+\abs{\la 
h_uh_d+\frac{\la^{\prime}}{2} s^2}^2+\abs{\la nh_d+\la_t\tilde{t}_L\tilde{t}_R}^2+\abs{\la nh_u}^2\\
\nonumber
&\phantom{space}&\hspace{0.2cm}+{\rm ~soft~susy~breaking~terms} + ...
\label{lag1}
\eea
where $n$ ($\tilde{n}$), $s$ ($\tilde{s}$), $h_u$ ($\tilde{h}_u$) and $h_d$ ($\tilde{h}_d$) are the scalar (fermionic) components of the superfields $N$, $S$, $H_u$ and $H_d$ 
respectively. 

To simplify the analysis, we make the reasonable assumption that the scalar tri-linear A-terms and bilinear B-terms are small and consequently we neglect their effects in cross sections. In particular we are setting the tri-linear A-terms, $A_{\la}=A_{\la^{\prime}}= 0$. We also neglect the D-term interactions as these give irrelevant
4-point Higgs interactions. 

To assess the viability of our dark matter candidate, we need to calculate its thermal relic abundance.   An important point to note is that the freeze-out temperature of our dark matter particle is higher than the electroweak phase transition temperature, $T_c$, for the range of dark matter masses $m_{\tilde{s}}\gsim 3 \tev$ we consider. (In our companion paper \cite{companion} we will explore the region of dark matter masses below $3\tev$.)  Consequently, in the relic density calculation, electroweak symmetry is still a good and no Higgs scalars will have vevs.  Moreover, above $T_c$, the fermionic states $\tilde{n}, \tilde{h}_u$ and $\tilde{h}_d$ are massless, as are all quarks and gauge bosons. The only massive fermionic state of interest is $\tilde{s}$ with mass $m_{\tilde s}$.  In the scalar sector, the thermally-corrected masses of the scalar $n$ states and MSSM Higgs bosons are taken to be negligible compared to $m_{\tilde s}$, which is a good approximation for the parameter range we are interested in.  

Taking $m_{\tilde{s}}\gsim 3 \tev$ does lead to one slight complication in our analysis in that the scalar state, $s$, has a very similar Boltzman factor compared to $\tilde{s}$ near the freeze-out temperature, $T_f$. This is due to the fact that the mass splitting between $s$ and  $\tilde{s}$ is small
\beq
m_s - m_{\tilde s} =
( m_{\tilde s}^2 + m_{\rm susy}^2 )^{1/2} - m_{\tilde s}  \simeq  m_{\rm susy}^2/m_{\tilde s}  <
T_{\rm f} \simeq m_{\tilde s}/25,
\eeq
where $m_{\rm susy}$ is the supersymmetry breaking scale, which is parametrically smaller than $m_{\tilde s}$. This means that the scalar $s$ and fermion ${\tilde s}$  states will freeze-out at roughly the same temperature and we have to consider the annihilation rates of the scalar states as well as the fermionic states\footnote{We remark in passing that our qualitative conclusions regarding the dark matter freeze out density would not be changed if a scalar component of $S$ were the lightest $Z_2$-odd state, say due to CP-violation. The Sommerfeld effect acts equally for both scalar and fermionic annihilating particles as explained in Section 3.}. 

In addition to the purely scalar interactions which follow from \eq{lag1} the fermionic interactions which are of importance in determining the relic abundance of our dark matter state are 
\bea
\nonumber
&&\hspace{-0.5cm} \frac{1}{\sqrt{2}}(\la \phi_n(\tilde{h}^0_{uM})^T C\tilde{h}^0_{dM}
+i\la a_n(\tilde{h}^0_{uM})^TC\ga_5\tilde{h}^0_{dM}
+\la\phi_u(\tilde{h}^0_{dM})^T C\tilde{n}_M
+i\la a_u(\tilde{h}^0_{dM})^T C\ga_5\tilde{n}_M)\\ \nonumber
&&+~\frac{1}{\sqrt{2}}(\la \phi_d(\tilde{h}^0_{uM})^T C\tilde{n}_M+i\la a_d(\tilde{h}^0_{uM})^T C\ga_5\tilde{n}_M
-\la^{\prime} \phi_s\tilde{s}^T_M C\tilde{n}_M
-i\la^{\prime} a_s\tilde{s}^T_M C\ga_5\tilde{n}_M)\\ \nonumber
&&-~\frac{1}{2\sqrt{2}}(\la^{\prime}\phi_n\tilde{s}^T_M C\tilde{s}_M+i\la^{\prime} a_n\tilde{s}^T_M C\ga_5\tilde{s}_M)\\ \nonumber
&&-~\frac{1}{\sqrt{2}}(\phi_n[\bar{\tilde{h}^-_{uD}}P_L\tilde{h}^-_{dD} + \bar{\tilde{h}^-_{dD}}P_R\tilde{h}^-_{uD}]
+ia_n[\bar{\tilde{h}^-_{uD}}P_L\tilde{h}^-_{dD} - \bar{\tilde{h}^-_{dD}}P_R\tilde{h}^-_{uD}])\\ \nonumber
&&-~h_d^-\bar{\tilde{h}^-_{uD}}P_L\tilde{n}_{M}-(h_d^-)^*\tilde{n}_{M}^TCP_R\tilde{h}^-_{uD}
-(h_u^+)^*\bar{\tilde{h}^-_{dD}}P_L\tilde{n}_{M}-h_d^+\tilde{n}_{M}^TCP_R\tilde{h}^-_{dD},
\eea
where $P_{L, R}=(1\pm\ga_5)/2$ and we have rewritten the fermionic states in terms of Majorana and Dirac spinors indicated by the subscripts $M$ and $D$ respectively. The scalar states have been written in terms of their CP-odd and CP-even components, denoted generically as $A_i=\frac{1}{\sqrt{2}}(\phi_i+ia_i)$, and $C$ is
the charge conjugation matrix. The subscripts $u$ and $d$ on the scalars refer to the Higgs ``up" and ``down" states.   

\section{The Sommerfeld Enhancement}
\label{som}

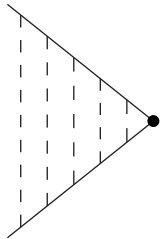
\begin{figure}[h]  
\begin{center}  
\begin{picture}(100,100)(-10,-10)  
\Line(50,40)(-5,-4)
\Line(50,40)(-5,84)
\DashLine(0,0)(0,80){5}
\DashLine(10,8)(10,72){5}
\DashLine(20,16)(20,64){5}
\DashLine(30,24)(30,56){5}
\DashLine(40,32)(40,48){5}
\GCirc(50,40){2}{0} 
\end{picture}  
\end{center}           
\caption{Generic Sommerfeld diagram. The ``blob" vertex represents all possible $S$-wave annihilations of the incoming states including s-channel, t-channel and annihilation via contact interactions.}  
\label{sommgen}  
\end{figure}

For dark matter particles moving at small relative velocities, the exchange of scalar states leads to an enhancement by factors depending on the inverse velocity, $1/v$. This Sommerfeld enhancement corresponds to the summation of a series of ladder diagrams where the scalar state is repeatedly exchanged (see Fig. \ref{sommgen}).  This enhancement is only significant if there exists an $S$-wave annihilation amplitude, otherwise the angular momentum barrier will suppress the effect.\footnote{If vector states are exchanged, there can either be an enhancement or suppression depending on the relative charges of the annihilating particles.} 

 
The calculation of the Sommerfeld enhancement can be formulated in terms of a non-relativistic quantum two-body problem with a potential acting between the
incoming particles. This is equivalent to the distorted Born-wave approximation common in nuclear physics. To a good 
approximation this leads to a dressing of the $S$-wave part of the tree-level cross sections with a multiplicative factor, 
\beq
\si=R\si^{\ell=0}_{tree }.
\eeq

The full calculation of $R$ can be involved and in many cases, including that of a Yukawa potential, cannot be solved analytically.   In our model the only particles which can act as the 
``rungs on the ladder" in the Sommerfeld diagram shown in Fig. \ref{sommgen} are the scalar $n$ states, the 
${\tilde n}$ fermions not contributing to the enhancement.  The non-relativistic potential which is relevant for all the diagrams we will consider is found to be
\beq
V=-\frac{\la^{\prime^2}}{8\pi r}e^{-m_nr},
\eeq
where $m_n$ is the mass of the particle acting as the ``rungs on the ladder". 
The Schr\"{o}dinger equation for the two dark matter particle state, $\psi$, with this potential reads
\beq
-\frac{1}{m_{\tilde s}}\frac{d^2\psi}{dr^2}+V.\psi=K\psi,
\eeq
where $K=Mv^2$ is the kinetic energy of the two dark matter particles in the center-of-mass frame, where each dark matter particle has velocity $v$. Using the outgoing boundary conditions,
$\psi^{\prime}(\infty)/\psi(\infty)=im_sv$,
R is given as $R=\abs{\psi(0)/\psi(\infty)}^2$. In the simple case we are considering, we can derive an analytic form for $R$. In the limit where the ratio $\ep\equiv
m_{n}/m_{\tilde s}=0$, $R$ takes the form~\cite{strumia},
\beq
R=\frac{y}{1-e^{-y}},
\eeq
where $y=\la^{\prime^2}/8v=\la^{\prime^2}/4v_r$ and $v_r=2v$ is the relative velocity between the two dark matter particles.
Taking the small $v_r$ limit we have
\beq
R\approx \frac{\la^{\prime^2}}{4v_{r}}
\eeq
and we see that this effect will be largest for small $v_r$.


\section{Calculation of the Relic Density}
\label{reliccalc}

We are now in a position to calculate the relic density of our dark matter candidate.  As mentioned in the previous section, in this paper we will restrict our analysis to dark matter particles with masses $m_{\tilde{s}}\ge 3 \tev$. Not only is this range of masses physically interesting, it also simplifies the analysis considerably due to the fact that freeze-out occurs at a temperature above the electroweak phase transition, thus leading to a situation in which no scalars have vevs. Consequently, the number of possible vertices contributing to the annihilation cross sections is reduced and the calculation of the relic abundance greatly simplified. 
  
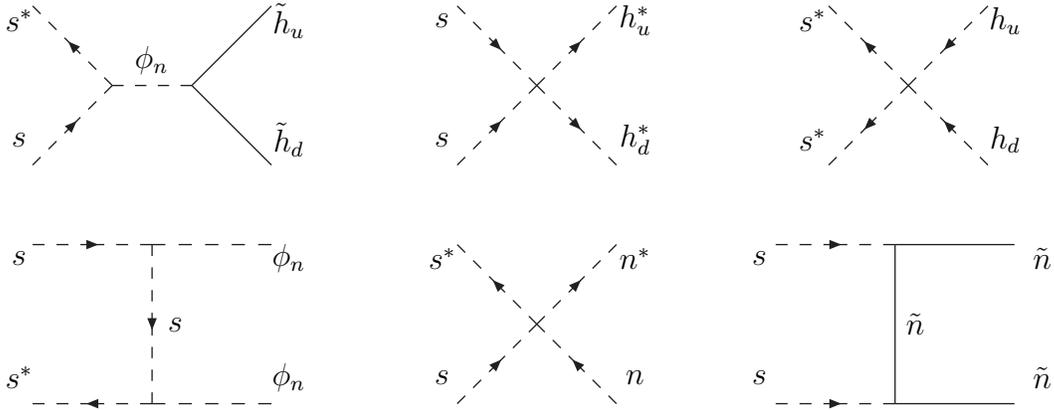
\begin{figure}[h!]  
\begin{center}  
\begin{picture}(520,180)(-10,-10)  
\DashArrowLine(30,90)(60,120) {5} 
\DashArrowLine(60,120)(30,150){5}
\DashLine(90,120)(60,120){5} 
\Line(90,120)(120,150)  
\Line(90,120)(120,90)
\Text(25,145)[]{$s^*$}  
\Text(25,100)[]{$s$}  
\Text(127,145)[]{$\tilde{h}_u$}  
\Text(127,100)[]{$\tilde{h}_d$}  
\Text(75,130)[]{$\phi_n$}  
\DashArrowLine(190,90)(220,120){5}   
\DashArrowLine(190,150)(220,120){5} 
\DashArrowLine(220,120)(250,150){5}  
\DashArrowLine(220,120)(250,90){5}
\Text(185,145)[]{$s$}  
\Text(185,100)[]{$s$}  
\Text(258,145)[]{$h_u^*$}  
\Text(258,100)[]{$h_d^*$}  
\DashArrowLine(360,120)(330,90){5}   
\DashArrowLine(360,120)(330,150){5} 
\DashArrowLine(390,150)(360,120){5}  
\DashArrowLine(390,90)(360,120){5}
\Text(325,145)[]{$s^*$}  
\Text(325,100)[]{$s^*$}  
\Text(398,145)[]{$h_u$}  
\Text(398,100)[]{$h_d$}  
\DashArrowLine(75,0)(30,0) {5} 
\DashLine(75,0)(120,0) {5} 
\DashArrowLine(30,60)(75,60){5}
\DashLine(75,60)(120,60){5}
\DashArrowLine(75,60)(75,0){5}
\Text(25,55)[]{$s$}  
\Text(25,10)[]{$s^*$}  
\Text(127,55)[]{$\phi_n$}  
\Text(127,10)[]{$\phi_n$}  
\Text(84,30)[]{$s$}  
\DashArrowLine(190,0)(220,30){5}   
\DashArrowLine(220,30)(190,60){5} 
\DashArrowLine(220,30)(250,60){5}  
\DashArrowLine(250,0)(220,30){5}
\Text(185,55)[]{$s^*$}  
\Text(185,10)[]{$s$}  
\Text(258,55)[]{$n^*$}  
\Text(258,10)[]{$n$}  
\DashArrowLine(310,0)(355,0) {5} 
\Line(355,0)(400,0)
\DashArrowLine(310,60)(355,60){5}
\Line(355,60)(400,60)
\Line(355,60)(355,0)
\Text(305,55)[]{$s$}  
\Text(305,10)[]{$s$}  
\Text(412,55)[]{$\tilde{n}$}  
\Text(412,10)[]{$\tilde{n}$}  
\Text(364,30)[]{$\tilde{n}$}  
\end{picture}  
\caption{Type I annihilation diagrams for the scalar $s$ states.} 
\label{scalarsann}  
\end{center}  
\end{figure}

There are three important types of diagram which determine the relic abundance of our dark matter particle. The first type (type I) involves the annihilation of two scalar $s$ states. We can have two scalar $s$ states annihilating into Higgs, Higgsinos, scalar $n$ states or fermionic $n$ states as depicted in Fig.~\ref{scalarsann}. 
For computational ease we take all states to be massless apart from $s$ and $\tilde{s}$ which have masses $m_{\tilde s}$ and
$m_{\tilde s}+ m_{\rm susy}^2/m_{\tilde s}$ respectively.

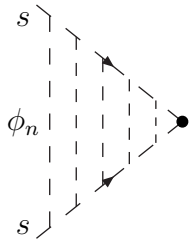
\begin{figure}[h!]  
\begin{center}  
\begin{picture}(100,100)(-10,-10)  
\DashArrowLine(-5,-4)(50,40){6}
\DashArrowLine(-5,84)(50,40){6}
\DashLine(0,0)(0,80){6} 
\DashLine(10,8)(10,72){5} 
\DashLine(20,16)(20,64){5} 
\DashLine(30,24)(30,56){5} 
\DashLine(40,32)(40,48){3} 
\GCirc(50,40){2}{0}
\Text(-10,79)[]{$s$}  
\Text(-10,0)[]{$s$}  
\Text(-10,40)[]{$\phi_n$}  
\end{picture}  
\end{center}           
\caption{Sommerfeld diagram for scalar annihilations.}  
\label{somms}  
\end{figure}

For all scalar annihilation diagrams we receive an enhancement from the Sommerfeld effect where the CP-even scalar, $\phi_n$, acts as the ``rungs on the ladder" between the annihilating scalar $s$ states as depicted in Fig.~\ref{somms} for the case of the annihilation of two $s$ states. The ``blob" vertex represents all possible ways of annihilating the $s$ states including s-channel, t-channel and annihilation via the 4-point vertex.

 
 The resulting self annihilation cross sections for the CP-even and CP-odd components of $s$ are
 \bea
 \si(\phi_s\phi_s\rightarrow[\tilde{h}_u\tilde{h}_d])&=&\si(a_sa_s\rightarrow[\tilde{h}_u\tilde{h}_d])=\frac{(\la^{\prime}\la)^2}{32\pi v_r m_{\tilde s}^2}\frac{y}{1-e^{-y}},\\
 \si(\phi_s\phi_s\rightarrow[h_uh_d])&=&\si(a_sa_s\rightarrow[h_uh_d])=\frac{(\la^{\prime}\la)^2}{32\pi v_r m_{\tilde s}^2}\frac{y}{1-e^{-y}},\\
 \si(a_s\phi_s\rightarrow(h_uh_d))&=&\frac{(\la^{\prime}\la)^2}{16\pi v_r m_{\tilde s}^2}\frac{y}{1-e^{-y}},\\
 \si(\phi_s\phi_s\rightarrow[nn])&=&\si(a_sa_s\rightarrow[nn])=\frac{\la^{\prime^4}}{32\pi v_r m_{\tilde s}^2}\frac{y}{1-e^{-y}},\\
 \si(\phi_s\phi_s\rightarrow \tilde{n}\tilde{n})&=&\si(a_sa_s\rightarrow \tilde{n}\tilde{n})=\frac{5\la^{\prime^4}}{32\pi v_r m_{\tilde s}^2}\frac{y}{1-e^{-y}},\\
 \si(\phi_sa_s\rightarrow \tilde{n}\tilde{n})&=&\frac{\la^{\prime^4}}{16\pi v_r m_{\tilde s}^2}\frac{y}{1-e^{-y}},
 \eea
where the factor $y/(1-e^{-y})$ accounts for the Sommerfeld enhancement and 

\bea
[\tilde{h}_u\tilde{h}_d]&=&\tilde{h}^0_u\tilde{h}^0_d, \tilde{h}^+_u\tilde{h}^-_d, (\tilde{h}^+_u\tilde{h}^-_d)^*,\nonumber \\ 
\left[h_uh_d\right]&=&\phi_u\phi_d, a_ua_d, h_u^+h_d^-, (h_u^+h_d^-)^*,\nonumber \\
(h_uh_d)&=&\phi_ua_d, a_u\phi_d, h_u^+h_d^-, (h_u^+h_d^-)^*,\nonumber \\ 
\left[nn\right]&=&\phi_n\phi_n, a_na_n,\nonumber \eea
represent all possible final states in each case.
 
\begin{figure}[h!]  
\begin{center}  
\begin{picture}(100,100)(-10,-10)  
\Line(50,40)(-5,-4)
\DashArrowLine(-5,84)(50,40){6}
\DashLine(0,0)(0,80){6} 
\DashLine(10,8)(10,72){5} 
\DashLine(20,16)(20,64){5} 
\DashLine(30,24)(30,56){5} 
\DashLine(40,32)(40,48){3}
\GCirc(50,40){2}{0}
\Text(-10,79)[]{$s$}  
\Text(-10,0)[]{$\tilde{s}$}  
\Text(-10,40)[]{$\phi_n$}  
\end{picture}  
\end{center}           
\caption{Sommerfeld diagrams for annihilation of $s$ and $\tilde{s}$.}  
\label{sommsstild}  
\end{figure}
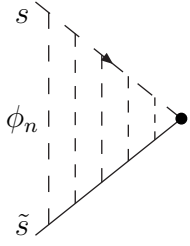

The second type of diagram (type II) we need to include is the annihilation of a scalar $s$ state with a fermionic $\tilde{s}$ state. The relevant diagrams are shown in Fig. \ref{scaferm}. Each process in Fig. \ref{scaferm} can also be enhanced by the Sommerfeld effect via the diagram shown in Fig. \ref{sommsstild}, where the ``blob" vertex represents both s-channel and t-channel processes. The enhancement factor is exactly the same in this case as it was for the annihilation of scalar $s$ states.


\begin{figure}[h!]  
\begin{center}  
\begin{picture}(520,90)(-10,-10)  
\Line(25,0)(55,30)
\DashArrowLine(25,60)(55,30){5}
\Line(85,30)(55,30)
\DashArrowLine(85,30)(115,60){5}
\Line(85,30)(115,0)
\Text(20,55)[]{$s$}  
\Text(20,10)[]{$\tilde{s}$}  
\Text(132,55)[]{$h_u, h_d$}  
\Text(132,10)[]{$\tilde{h}_d, \tilde{h}_u$}  
\Text(70,40)[]{$\tilde{n}$}  
\Line(230,0)(175,0) 
\Line(230,0)(265,0)
\DashArrowLine(175,60)(220,60){5}
\DashLine(230,60)(265,60){5}
\DashArrowLine(220,60)(220,0){5}
\Text(170,55)[]{$s$}  
\Text(170,10)[]{$\tilde{s}$}  
\Text(272,55)[]{$\phi_n$}  
\Text(272,10)[]{$\tilde{n}$}  
\Text(226,30)[]{$s$}  
\Line(310,0)(355,0)
\DashLine(355,0)(400,0) {5} 
\DashArrowLine(310,60)(355,60){5}
\Line(355,60)(400,60)
\Line(355,60)(355,0)
\Text(305,55)[]{$s$}  
\Text(305,10)[]{$\tilde{s}$}  
\Text(412,55)[]{$\tilde{n}$}  
\Text(412,10)[]{$a_{n}$}  
\Text(364,30)[]{$\tilde{s}$}  
\end{picture}  
\end{center}  
\caption{Type II: Annihilation of $s$ with $\tilde{s}$.}  
\label{scaferm}  
\end{figure}
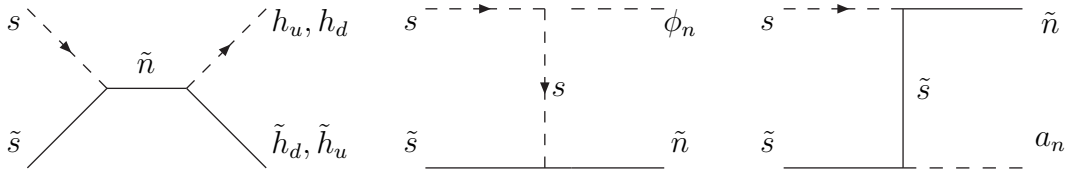  

 
 The cross sections for these processes are found to be
 \bea
\si(\phi_s\tilde{s}\rightarrow (h_i\tilde{h}_j))&=&\si(a_s\tilde{s}\rightarrow (h_i\tilde{h}_j))=\frac{(\la^{\prime}\la)^2}{32\pi v_r m_{\tilde s}^2}\frac{y}{1-e^{-y}},\\
\si(\phi_s\tilde{s}\rightarrow (n\tilde{n}))&=&\si(a_s\tilde{s}\rightarrow (n\tilde{n}))=\frac{(\la^{\prime})^4}{32\pi v_r m_{\tilde s}^2}\frac{y}{1-e^{-y}}
 \eea where 
 \bea
 (h_i\tilde{h}_j)&=&\phi_d\tilde{h}_u^0, a_d\tilde{h}_u^0, h^-_d\tilde{h}_u^+, (h^-_d\tilde{h}_u^+)^*, \phi_u\tilde{h}^0_d, a_u\tilde{h}_d^0, h^+_u\tilde{h}_d^-, (h^+_u\tilde{h}^-_d)^*, \nonumber \\ (n\tilde{n})&=&a_n\tilde{n}, \phi_n\tilde{n},\nonumber
 \eea
represent all possible final states for each process.

\begin{figure}[h]  
\begin{center}  
\begin{picture}(520,90)(-10,-10)  
\Line(70,0)(100,30)
\Line(70,60)(100,30)
\DashLine(130,30)(100,30){5}
\Line(130,30)(160,60)
\Line(130,30)(160,0)
\Text(65,55)[]{$\tilde{s}$}  
\Text(65,10)[]{$\tilde{s}$}  
\Text(168,55)[]{$\tilde{h}_u$}  
\Text(168,10)[]{$\tilde{h}_d$}  
\Text(115,40)[]{$a_n$}  
\Line(250,0)(295,0)
\DashLine(295,0)(340,0){5} 
\Line(250,60)(295,60)
\DashLine(295,60)(340,60){5}
\Line(295,60)(295,0)
\Text(245,55)[]{$ \tilde{s}$}  
\Text(245,10)[]{$\tilde{s}$}  
\Text(352,55)[]{$\phi_n$}  
\Text(352,10)[]{$a_{n}$}  
\Text(304,30)[]{$\tilde{s}$} 
\end{picture}  
\end{center}  
\caption{Type III: Annihilation of two $\tilde{s}$ states.}  
\label{stild}  
\end{figure}
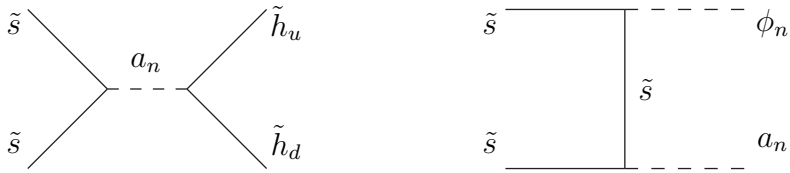

The third type of process (type III) is the annihilation of the $\tilde{s}$ states. In the electroweak symmetric limit, there are no vevs for the $n$ or Higgs states, neither is there a
tri-linear scalar $A$ term,  $A_{\la}nh_uh_d$ as we have approximated this term to be zero. This means that the only Higgs-like final states from $\tilde{s}$ annihilation will be products of neutralinos and charginos. We can also have t-channel exchange of a $\tilde{s}$, which produces a CP-odd final state pair, $a_n\phi_n$. All possible diagrams with non-zero $S$-wave amplitudes are shown in Fig. \ref{stild}.

\begin{figure}[h]  
\begin{center}  
\begin{picture}(100,100)(-10,-10)  
\Line(-5,84)(50,40)
\Line(-5,-4)(50,40) 
\DashLine(0,0)(0,80){6} 
\DashLine(10,8)(10,70){6}
\DashLine(20,16)(20,64){5} 
\DashLine(30,24)(30,56){5}
\DashLine(40,32)(40,48){3} 
\GCirc(50,40){2}{0}
\Text(-10,79)[]{$\tilde{s}$}  
\Text(-10,0)[]{$\tilde{s}$}  
\Text(-10,40)[]{$\phi_n$}  
\end{picture}  
\end{center}  
\caption{Sommerfeld diagrams for the annihilation of two $\tilde{s}$ states.}  
\label{stildsom}  
\end{figure}
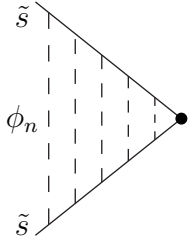

The $\tilde{s}\tilde{s}$ annihilation processes can be enhanced by the Sommerfeld factor if the initial $\tilde{s}$ pair are in an $S$-wave state. The corresponding Sommerfeld diagram for $\tilde{s}\tilde{s}$ annihilation is shown in Fig. \ref{stildsom}, where the ``blob" vertex represents both s-channel and t-channel processes.


The resulting cross sections for the $\tilde{s}$ annihilations are found to be
\bea
\si(\tilde{s}\tilde{s}\rightarrow[\tilde{h}_u\tilde{h}_d])&=&\frac{(\la^{\prime}\la)^2}{64\pi v_r m_{\tilde s}^2}\frac{y}{1-e^{-y}},\\
\si(\tilde{s}\tilde{s}\rightarrow\tilde{n}\tilde{n})&=&\frac{(\la^{\prime})^4}{64\pi v_r m_{\tilde s}^2}\frac{y}{1-e^{-y}}
\eea
where $[\tilde{h}_u\tilde{h}_d]$ was defined earlier in this section.

We now have all of the cross sections needed to determine the relic density.  As we have two states freezing out almost simultaneously (our dark matter state 
$\tilde{s}$ and its scalar partner $s$) we must be careful to include the effects of the heavier state in the  calculation of the relic abundance of the dark matter species. We follow Refs.~\cite{GS,GG} in calculating the final relic abundance of our dark matter candidate.

If we relabel our two states, $\tilde{s}$ and $s$, as $s_1$ and $s_2$ respectively, the type of reaction that will determine the freeze-out of our two particles is
\beq
\si_{ij}=\si(s_is_j\rightarrow XX^{\prime}),
\eeq
where $X$ and $X^{\prime}$ will be some combination of Higgses, higgsinos, fermionic $\tilde{n}$ states and scalar $n$ states, which will decay to lighter MSSM degrees of freedom. Taking into account all possible diagrams, the three cross sections we are concerned with have the following forms
\bea
\si(s_1s_1\rightarrow XX^{\prime})&=&\frac{[(\la^{\prime}\la)^2+(\la^{\prime})^4]}{64\pi v_r m_{\tilde s}^2}\frac{y}{1-e^{-y}},\\
\si(s_1s_2 \rightarrow XX^{\prime})&=&\frac{[(\la^{\prime}\la)^2+(\la^{\prime})^4]}{32\pi v_r m_{\tilde s}^2}\frac{y}{1-e^{-y}}, \\ 
\si(s_2s_2 \rightarrow XX^{\prime})&=&\frac{[3(\la^{\prime}\la)^2+7(\la^{\prime})^4]}{64\pi v_r m_{\tilde s}^2}\frac{y}{1-e^{-y}},
\eea
where we have averaged over the components of the initial scalar states where appropriate.

We assume that any $s_2$ states remaining after freeze-out will eventually decay down to $s_1XX^{\prime}$. This means that the total number density of our dark matter particle 
will be equal to the sum of the $s_1$ and $s_2$ number densities at freeze-out.

In order to calculate the relic density we define the following useful quantities \cite{GS}
\beq
r_i\equiv \frac{g_i(1+\Delta_i)^{3/2}\exp[-x\Delta_i]}{g_{\rm eff}},
\eeq
where
\beq
\Delta_i=(m_i-m_1)/m_1, 
\eeq
and 
\beq
g_{\rm eff}=\sum_{i=1}^2 g_i(1+\Delta_i)^{3/2}\exp[-x\Delta_i],
\eeq
where $g_i$ is the number of degrees of freedom of $s_i$, $m_i$ is the mass of $s_i$ and $x=m_{\tilde{s}}/T$. Of course in our case we only have two different species of 
particle and so only $\Delta_2$ is non-zero. In fact as $s_1=\tilde{s}$ and $s_2=s$, we have $\Delta_2= m_s - m_{\tilde s}  \simeq  m_{\rm susy}^2/m_{\tilde s}$. Each of our 
$s_i$ states have $g_i=2$ degrees of freedom. Following Ref.~\cite{GS}, we find the freeze-out temperature, $T_f$, by iteratively  solving the equation
\beq
x_f=\ln\left[\frac{0.038g_{\rm eff}M_{pl}m_{\tilde{s}}\vev{\si_{eff}v_{r}}}{g_{\star}^{1/2}x_f^{1/2}}\right],
\eeq  
where $x_f=m_{\tilde s}/T_f$ and 
\beq
\si_{\rm eff}=\sum_{i,j}^{2}\si_{ij}r_ir_j=\sum_{i,j}^{2}\si_{ij}\frac{g_ig_j}{g^2_{\rm eff}}(1+\Delta_i)^{3/2}(1+\Delta_j)^{3/2}\exp(-x(\Delta_i+\Delta_j)).
\eeq

The final relic density is given by~\cite{GS}
\beq
\Omega h^2=\frac{1.07\times 10^9x_f}{g_{\star}^{1/2}M_{pl}(\gev)J}, \; \mbox{where} \; J=\int_{x_f}^{\infty}x^{-2}a_{\rm eff}dx
\eeq
and $g_{\star}$ is the total number of relativistic degrees of freedom at $T_{f}$. In our calculation of the relic density we will take $g_{\star}=248$, which includes all MSSM degrees of freedom plus the four associated with the extra superfield $N$. In order for this to be correct the masses of all these states must be below $T_{f}\sim m_{\tilde{s}}/25$, which will be true when we take $m_{\tilde{s}}\ge 3\tev$ and $m_{\rm susy}=100\gev$ as an example parameter set.

\begin{figure}[t]
\centering\leavevmode
\includegraphics[width=0.9\textwidth]{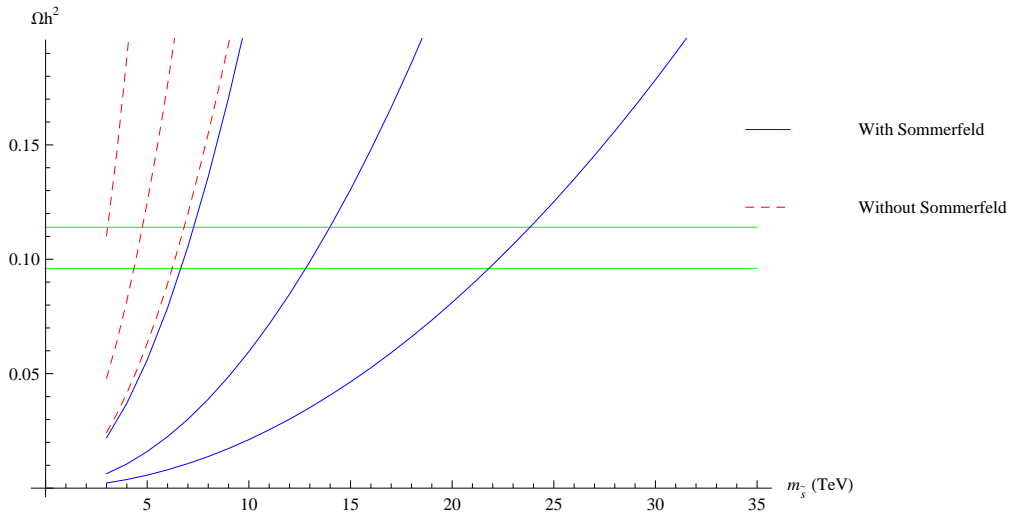}
\caption{$\Omega h^2$ as function of $m_{\tilde{s}}$ at fixed $\la$ and $\la^{\prime}$. The red dashed lines correspond to the case where the Sommerfeld correction is not 
included where as the blue solid lines correspond to the case when it is included. The furthest most left line for each colour corresponds to when $\la=\la^{\prime}=2$, the
middle lines are when $\la=\la^{\prime}=2.5$ and the lines furthest right are when $\la=\la^{\prime}=3$. All plots are produced using $m_{\rm susy}=100\gev$.}
\label{omms}
\end{figure}

It is instructive to compare the two cases of when we correctly include the Sommerfeld factor in cross sections and when this contribution is absent. The comparison is most clear 
when we plot the relic density, $\Omega_{\tilde{s}} h^2$, against $m_{\tilde{s}}$ as shown in Fig.~\ref{omms}. In Fig.~\ref{omms}, the red dashed lines correspond to the case where the 
Sommerfeld factor (given by $R=y/(1-e^{-y})$) is not included in the cross sections, while the blue solid lines correspond to the case where it is. The three lines for each case (with 
and without the Sommerfeld effect), starting from the furthest left, correspond to $\la=\la^{\prime}=$2, 2.5, 3 respectively.  The two lines parallel with the $m_{\tilde{s}}$ axis 
correspond to the WMAP allowed range for the dark matter relic abundance, inferred from the combination of $\Omega_M h^2 = 0.1277^{+0.0080}_{-0.0079}$ and $\Omega_b h^2=0.02229 \pm 0.00073$ \cite{WMAP}.

For each line (of fixed coupling), the relic density increases as we increase the mass, $m_{\tilde{s}}$, as we would expect. Comparing sets of contours with the same couplings ($\la=\la^{\prime}$), we see the dramatic effect of the Sommerfeld enhancement. When the Sommerfeld enhancement is included, the annihilation cross sections are increased, thus depleting the number density of the dark matter particles which survive after freeze-out. The bottom line is that the Sommerfeld enhancement allows for very heavy dark matter particles to provide the required dark matter relic abundance. From Fig.~\ref{omms} we can see that the maximum mass consistent with the WMAP allowed range when we have $\la=\la^{\prime}=$3 is close to $25\tev$.

\begin{figure}[t]
\centering\leavevmode
\includegraphics[width=0.9\textwidth,height=0.55\textwidth]{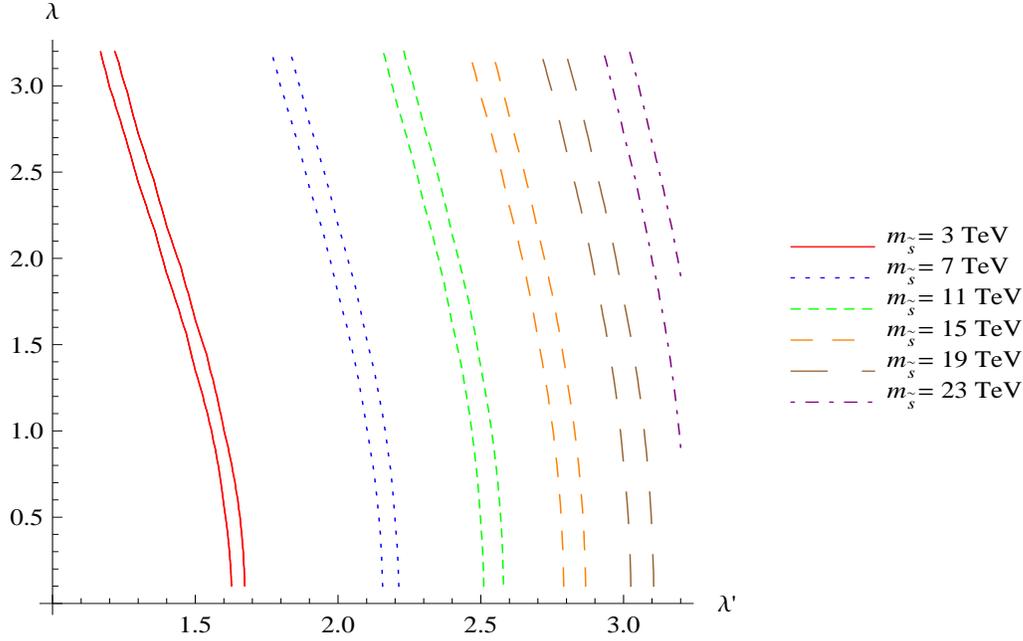}
\caption{Plots of pairs of contours for the allowed range of $\Omega h^2$ in the $\la-\la^{\prime}$ parameter plane for different values of the mass, $m_{\tilde{s}}$. We have 
contours corresponding to masses $3, 7, 11, 15, 19, 23\tev$. The contours are produced using $m_{\rm susy}=100\gev$.}
\label{lala}
\end{figure}

The results of a full numerical scan (including the Sommerfeld enhancement) over the three parameters $\la, \la^{\prime}$ and $m_{\tilde{s}}$ is shown in Fig. \ref{lala}. Here, we plot contours corresponding to the allowed range of $\Omega_{\tilde{s}} h^2$ in the $\la-\la^{\prime}$ plane. Each pair of contours correspond to a different value of the mass, $m_{\tilde{s}}$, between 3 and 23 TeV. The left (right) contour of each pair corresponds to the higher (lower) end of the allowed range in $\Omega_{\tilde{s}} h^2$. 

Although we show contours only for discrete choices of $m_{\tilde{s}}$, the remaining regions of the $\la-\la^{\prime}$ plane are filled for intermediate values of the dark matter mass.\footnote{There will be an upper limit on how large the couplings can be, which is determined by insisting we have perturbativity up to our cut off scale.  As shown in Section~\ref{fathiggs}, large couplings of the size considered here are shown to be natural in a consistent UV completion.} The effect of the Sommerfeld enhancement is to pull the pairs of contours downward towards the bottom left corner of the $\la-\la^{\prime}$ plane. This allows us to have the correct relic density for a given dark matter mass for smaller values of the couplings. 

On examination of the parameter region shown in Fig. \ref{lala}, it is interesting and perhaps important to note that we are able to generate the correct dark matter relic density  (for a given mass) for relatively small $\la$ couplings, provided we have a large enough $\la^{\prime}$ coupling. The main reason for this is that only the $\la^{\prime}$ coupling appears in the Sommerfeld enhancement and as we can see from Fig.~\ref{omms}, it is the Sommerfeld enhancement that allows us to have large masses for the dark matter particles. With this in mind, it is also interesting to note that as we have annihilation diagrams which depend on $\la^{\prime}$ only, we can take $\la$ to be small (around 0.1 for example) and still have viable dark matter with masses up to around $23\tev$.

\section{Direct and Indirect Detection}
\label{detection}

Although we intend to discuss the prospects for the direct and indirect detection of heavy Higgs Portal dark matter in some detail in a companion paper to follow~\cite{companion} we will here briefly touch upon this subject.  We find that the direct detection phenomenology is fairly conventional and, although present experiments do not yet lead to restrictive limits, a sizeable fraction of the expected parameter space will be covered by proposed next generation detectors.  In contrast, the indirect signals are greatly modified by the potentially very large Sommerfeld enhancements. 

\subsection{Direct Detection}

Experimental programs designed to observe the elastic scattering of WIMPs with nuclei are collectively known as direct detection. The dark matter particles
in our model, $\tilde{s}$, interact with quarks in nuclei through the effective scalar interaction given by
\beq
\mathcal{L}=\sum_{U=u,c,t}C_U\tilde{s}\tilde{s}\, \bar{U}U+\sum_{D=d,s,b}C_D\tilde{s}\tilde{s}\, \bar{D}D,
\eeq
where
\beq
C_U=\sum_i\frac{\la_UV_{1i}V_{2i}\la^{\prime}}{2m_{h_i}^2}~ {\rm and}~
C_D=\sum_i\frac{\la_DV_{1i}V_{3i}\la^{\prime}}{2m_{h_i}^2},
\eeq
and the mixing matrix $V_{ij}$ specifies the admixture of $n$, $h_u^0$, and $h_d^0$ states in the neutral scalar mass eigenstates, $h_i$, with lightest neutral Higgs state being denoted
$h_1$. Unlike in the case of many other dark matter candidates, there is no contribution from $Z$ exchange in this model. Note that the kinematics of the interaction (even if we consider scattering of individual nucleons or even quarks with the dark matter) are such that we are outside of the range for which the Sommerfeld enhancement is important. 

Following Refs.~\cite{direct,dmreview}, we estimate that this interaction leads to an elastic scattering cross secton per nucleon of
\beq
\si_{\tilde{s}N} \sim 2 \times 10^{-7} \,\rm{pb}\, \bigg(\frac{V_{ij}}{0.5}\bigg)^4\,\, \bigg(\frac{\la^{\prime}}{3}\bigg)^2 \, \left(\frac{120\gev}{m_{h_1}}\right)^4.
\eeq

For the range of masses we are interested in here, this cross section is below the current constraints from experiments such as XENON~\cite{xenon} and CDMS~\cite{cdms}, but is likely to be reached in the next few years. For less optimal values of $\lambda{^\prime}$, $m_{h_1}$ or $V_{ij}$, however, the prospects for direct detection could be considerably more difficult.

\subsection{Indirect Detection}

In addition to direct searches for dark matter, astronomers are also searching for the products of dark matter annihilations, including gamma-rays, neutrinos, positrons and antiprotons~\cite{dmreview2}. These efforts are known as indirect detection.

The dark matter annihilation rate, and thus indirect detection rates, can be enormously enhanced due to the Sommerfeld effect.  Depending on the astrophysical environment being considered, annihilation rates can be enhanced by factors of $10^3$ to $10^5$ or even greater due to the slow relative velocities of dark matter particles.  In fact, the velocity dependence of the enhancement factor can potentially favour such astrophysical objects as dwarf satellite galaxies of the Milky Way (due to the extremely low velocity dispersion) as sites for indirect detection, rather than the
central regions of the Milky Way itself.   A full calculation of the expected flux depends upon a detailed knowledge both of the
resonance structure of the Sommerfeld enhancement in the non-coulombic and low $v_r$ regime and
of the sizes of vacuum expectation values and interaction terms in the scalar $(S,N)$-Higgs sector.
A preliminary estimate shows that current indirect detection experiments do not impose a useful limit
on heavy Higgs portal dark matter, but that there is a potential for significant signals in future observations~\cite{companion}.


\section{UV Completion as a Fat Higgs Model}
\label{fathiggs}

It should be noted that the sizes of the couplings we have taken in the analysis of section \ref{reliccalc} are the values for typical momentum transfers at freeze-out ($\sim m_{\tilde{s}}\be_{fo}$, with $\be_{fo}\sim 0.2$). The cut-off of our effective theory will be related to the energy scale at which our couplings become non-perturbative, which for definiteness we take to be where the two loop terms in the renormalisation group equations for $\la$ and $\la^{\prime}$ become of order the one loop terms.  For example, if we take $m_{\tilde{s}}=3\tev$, $\la=0.8$ and $\la^{\prime}=1.6$ the cut-off is $\sim4000\tev$. A more extreme example is where we take $m_{\tilde{s}}=23\tev$, $\la=1.0$ and $\la^{\prime}=3.2$ with cut-off $\sim70\tev$. The consequence of having a cut-off below the GUT scale is that we are motivated to think about how this model can be UV completed. We emphasise that for the analysis of the dark matter properties and thermal freeze out the effective low energy lagrangian, \eq{sup1}, is appropriate as there is a large separation between $m_{\tilde{s}}\be_{fo}$ and the cut-off even for the most extreme case we consider, $m_{\tilde{s}}=23\tev$.

One possible way to UV complete our model and justify the choice of large couplings $\la, \la^{\prime}$ is to have some strongly interacting physics which dynamically generates the superpotential $S$ mass. It is noteworthy that the ``Fat Higgs model" of  Ref.~\cite{Harnik:2003rs} provides exactly such a UV completion.  With this in mind, we will now describe how our effective theory can arise in a certain limit of the Fat Higgs models.

The Fat Higgs model is an $N=1$ supersymmetric $SU(2)$ gauge theory with six doublets with the quantum numbers shown in Table \ref{tab1}.
\begin{table}\centering
\begin{tabular}{|l|llllll|}
\cline{1-7}
\vbox to1.63ex{\vspace{1pt}\vfil\hbox to10ex{\hfil Superfield\hfil}} & 
\vbox to1.63ex{\vspace{1pt}\vfil\hbox to8ex{\hfil $SU(2)_L$\hfil}} & 
\vbox to1.63ex{\vspace{1pt}\vfil\hbox to8ex{\hfil $SU(2)_H$\hfil}} & 
\vbox to1.63ex{\vspace{1pt}\vfil\hbox to8ex{\hfil $SU(2)_R$\hfil}} & 
\vbox to1.63ex{\vspace{1pt}\vfil\hbox to8ex{\hfil $SU(2)_g$\hfil}} & 
\vbox to1.63ex{\vspace{1pt}\vfil\hbox to8ex{\hfil $U(1)_R$\hfil}} & 
\vbox to1.63ex{\vspace{1pt}\vfil\hbox to8ex{\hfil $Z_2$\hfil}} \\

\cline{1-7}
\vbox to1.63ex{\vspace{1pt}\vfil\hbox to10ex{\hfil $T_1$\hfil}} & 
\vbox to1.63ex{\vspace{1pt}\vfil\hbox to8ex{\hfil 2\hfil}} & 
\vbox to1.63ex{\vspace{1pt}\vfil\hbox to8ex{\hfil 2\hfil}} & 
\vbox to1.63ex{\vspace{1pt}\vfil\hbox to8ex{\hfil 1\hfil}} & 
\vbox to1.63ex{\vspace{1pt}\vfil\hbox to8ex{\hfil 1\hfil}} & 
\vbox to1.63ex{\vspace{1pt}\vfil\hbox to8ex{\hfil 0\hfil}} & 
\vbox to1.63ex{\vspace{1pt}\vfil\hbox to8ex{\hfil +\hfil}} \\


\vbox to1.63ex{\vspace{1pt}\vfil\hbox to10ex{\hfil $T_2$\hfil}} & 
\vbox to1.63ex{\vspace{1pt}\vfil\hbox to8ex{\hfil 2\hfil}} & 
\vbox to1.63ex{\vspace{1pt}\vfil\hbox to8ex{\hfil 2\hfil}} & 
\vbox to1.63ex{\vspace{1pt}\vfil\hbox to8ex{\hfil 1\hfil}} & 
\vbox to1.63ex{\vspace{1pt}\vfil\hbox to8ex{\hfil 1\hfil}} & 
\vbox to1.63ex{\vspace{1pt}\vfil\hbox to8ex{\hfil 0\hfil}} & 
\vbox to1.63ex{\vspace{1pt}\vfil\hbox to8ex{\hfil -\hfil}} \\


\vbox to1.63ex{\vspace{1pt}\vfil\hbox to10ex{\hfil $T_3$\hfil}} & 
\vbox to1.63ex{\vspace{1pt}\vfil\hbox to8ex{\hfil 1\hfil}} & 
\vbox to1.63ex{\vspace{1pt}\vfil\hbox to8ex{\hfil 2\hfil}} & 
\vbox to1.63ex{\vspace{1pt}\vfil\hbox to8ex{\hfil 2\hfil}} & 
\vbox to1.63ex{\vspace{1pt}\vfil\hbox to8ex{\hfil 1\hfil}} & 
\vbox to1.63ex{\vspace{1pt}\vfil\hbox to8ex{\hfil 1\hfil}} & 
\vbox to1.63ex{\vspace{1pt}\vfil\hbox to8ex{\hfil -\hfil}} \\


\vbox to1.63ex{\vspace{1pt}\vfil\hbox to10ex{\hfil $T_4$\hfil}} & 
\vbox to1.63ex{\vspace{1pt}\vfil\hbox to8ex{\hfil 1\hfil}} & 
\vbox to1.63ex{\vspace{1pt}\vfil\hbox to8ex{\hfil 2\hfil}} & 
\vbox to1.63ex{\vspace{1pt}\vfil\hbox to8ex{\hfil 2\hfil}} & 
\vbox to1.63ex{\vspace{1pt}\vfil\hbox to8ex{\hfil 1\hfil}} & 
\vbox to1.63ex{\vspace{1pt}\vfil\hbox to8ex{\hfil 1\hfil}} & 
\vbox to1.63ex{\vspace{1pt}\vfil\hbox to8ex{\hfil +\hfil}} \\


\vbox to1.63ex{\vspace{1pt}\vfil\hbox to10ex{\hfil $T_5$\hfil}} & 
\vbox to1.63ex{\vspace{1pt}\vfil\hbox to8ex{\hfil 1\hfil}} & 
\vbox to1.63ex{\vspace{1pt}\vfil\hbox to8ex{\hfil 2\hfil}} & 
\vbox to1.63ex{\vspace{1pt}\vfil\hbox to8ex{\hfil 1\hfil}} & 
\vbox to1.63ex{\vspace{1pt}\vfil\hbox to8ex{\hfil 2\hfil}} & 
\vbox to1.63ex{\vspace{1pt}\vfil\hbox to8ex{\hfil 1\hfil}} & 
\vbox to1.63ex{\vspace{1pt}\vfil\hbox to8ex{\hfil +\hfil}} \\


\vbox to1.63ex{\vspace{1pt}\vfil\hbox to10ex{\hfil $T_6$\hfil}} & 
\vbox to1.63ex{\vspace{1pt}\vfil\hbox to8ex{\hfil 1\hfil}} & 
\vbox to1.63ex{\vspace{1pt}\vfil\hbox to8ex{\hfil 2\hfil}} & 
\vbox to1.63ex{\vspace{1pt}\vfil\hbox to8ex{\hfil 1\hfil}} & 
\vbox to1.63ex{\vspace{1pt}\vfil\hbox to8ex{\hfil 2\hfil}} & 
\vbox to1.63ex{\vspace{1pt}\vfil\hbox to8ex{\hfil 1\hfil}} & 
\vbox to1.63ex{\vspace{1pt}\vfil\hbox to8ex{\hfil +\hfil}} \\

\cline{1-7}
\vbox to1.63ex{\vspace{1pt}\vfil\hbox to10ex{\hfil $P_{11}$\hfil}} & 
\vbox to1.63ex{\vspace{1pt}\vfil\hbox to8ex{\hfil 2\hfil}} & 
\vbox to1.63ex{\vspace{1pt}\vfil\hbox to8ex{\hfil 1\hfil}} & 
\vbox to1.63ex{\vspace{1pt}\vfil\hbox to8ex{\hfil 1\hfil}} & 
\vbox to1.63ex{\vspace{1pt}\vfil\hbox to8ex{\hfil 2\hfil}} & 
\vbox to1.63ex{\vspace{1pt}\vfil\hbox to8ex{\hfil 1\hfil}} & 
\vbox to1.63ex{\vspace{1pt}\vfil\hbox to8ex{\hfil +\hfil}} \\


\vbox to1.63ex{\vspace{1pt}\vfil\hbox to10ex{\hfil $P_{12}$\hfil}} & 
\vbox to1.63ex{\vspace{1pt}\vfil\hbox to8ex{\hfil 2\hfil}} & 
\vbox to1.63ex{\vspace{1pt}\vfil\hbox to8ex{\hfil 1\hfil}} & 
\vbox to1.63ex{\vspace{1pt}\vfil\hbox to8ex{\hfil 1\hfil}} & 
\vbox to1.63ex{\vspace{1pt}\vfil\hbox to8ex{\hfil 2\hfil}} & 
\vbox to1.63ex{\vspace{1pt}\vfil\hbox to8ex{\hfil 1\hfil}} & 
\vbox to1.63ex{\vspace{1pt}\vfil\hbox to8ex{\hfil +\hfil}} \\


\vbox to1.63ex{\vspace{1pt}\vfil\hbox to10ex{\hfil $P_{21}$\hfil}} & 
\vbox to1.63ex{\vspace{1pt}\vfil\hbox to8ex{\hfil 2\hfil}} & 
\vbox to1.63ex{\vspace{1pt}\vfil\hbox to8ex{\hfil 1\hfil}} & 
\vbox to1.63ex{\vspace{1pt}\vfil\hbox to8ex{\hfil 1\hfil}} & 
\vbox to1.63ex{\vspace{1pt}\vfil\hbox to8ex{\hfil 2\hfil}} & 
\vbox to1.63ex{\vspace{1pt}\vfil\hbox to8ex{\hfil 1\hfil}} & 
\vbox to1.63ex{\vspace{1pt}\vfil\hbox to8ex{\hfil -\hfil}} \\


\vbox to1.63ex{\vspace{1pt}\vfil\hbox to10ex{\hfil $P_{22}$\hfil}} & 
\vbox to1.63ex{\vspace{1pt}\vfil\hbox to8ex{\hfil 2\hfil}} & 
\vbox to1.63ex{\vspace{1pt}\vfil\hbox to8ex{\hfil 1\hfil}} & 
\vbox to1.63ex{\vspace{1pt}\vfil\hbox to8ex{\hfil 1\hfil}} & 
\vbox to1.63ex{\vspace{1pt}\vfil\hbox to8ex{\hfil 2\hfil}} & 
\vbox to1.63ex{\vspace{1pt}\vfil\hbox to8ex{\hfil 1\hfil}} & 
\vbox to1.63ex{\vspace{1pt}\vfil\hbox to8ex{\hfil -\hfil}} \\


\vbox to1.63ex{\vspace{1pt}\vfil\hbox to10ex{\hfil $Q_{11}$\hfil}} & 
\vbox to1.63ex{\vspace{1pt}\vfil\hbox to8ex{\hfil 1\hfil}} & 
\vbox to1.63ex{\vspace{1pt}\vfil\hbox to8ex{\hfil 1\hfil}} & 
\vbox to1.63ex{\vspace{1pt}\vfil\hbox to8ex{\hfil 2\hfil}} & 
\vbox to1.63ex{\vspace{1pt}\vfil\hbox to8ex{\hfil 2\hfil}} & 
\vbox to1.63ex{\vspace{1pt}\vfil\hbox to8ex{\hfil 1\hfil}} & 
\vbox to1.63ex{\vspace{1pt}\vfil\hbox to8ex{\hfil -\hfil}} \\


\vbox to1.63ex{\vspace{1pt}\vfil\hbox to10ex{\hfil $Q_{12}$\hfil}} & 
\vbox to1.63ex{\vspace{1pt}\vfil\hbox to8ex{\hfil 1\hfil}} & 
\vbox to1.63ex{\vspace{1pt}\vfil\hbox to8ex{\hfil 1\hfil}} & 
\vbox to1.63ex{\vspace{1pt}\vfil\hbox to8ex{\hfil 2\hfil}} & 
\vbox to1.63ex{\vspace{1pt}\vfil\hbox to8ex{\hfil 2\hfil}} & 
\vbox to1.63ex{\vspace{1pt}\vfil\hbox to8ex{\hfil 1\hfil}} & 
\vbox to1.63ex{\vspace{1pt}\vfil\hbox to8ex{\hfil -\hfil}} \\


\vbox to1.63ex{\vspace{1pt}\vfil\hbox to10ex{\hfil $Q_{21}$\hfil}} & 
\vbox to1.63ex{\vspace{1pt}\vfil\hbox to8ex{\hfil 1\hfil}} & 
\vbox to1.63ex{\vspace{1pt}\vfil\hbox to8ex{\hfil 1\hfil}} & 
\vbox to1.63ex{\vspace{1pt}\vfil\hbox to8ex{\hfil 2\hfil}} & 
\vbox to1.63ex{\vspace{1pt}\vfil\hbox to8ex{\hfil 2\hfil}} & 
\vbox to1.63ex{\vspace{1pt}\vfil\hbox to8ex{\hfil 1\hfil}} & 
\vbox to1.63ex{\vspace{1pt}\vfil\hbox to8ex{\hfil +\hfil}} \\


\vbox to1.63ex{\vspace{1pt}\vfil\hbox to10ex{\hfil $Q_{22}$\hfil}} & 
\vbox to1.63ex{\vspace{1pt}\vfil\hbox to8ex{\hfil 1\hfil}} & 
\vbox to1.63ex{\vspace{1pt}\vfil\hbox to8ex{\hfil 1\hfil}} & 
\vbox to1.63ex{\vspace{1pt}\vfil\hbox to8ex{\hfil 2\hfil}} & 
\vbox to1.63ex{\vspace{1pt}\vfil\hbox to8ex{\hfil 2\hfil}} & 
\vbox to1.63ex{\vspace{1pt}\vfil\hbox to8ex{\hfil 1\hfil}} & 
\vbox to1.63ex{\vspace{1pt}\vfil\hbox to8ex{\hfil +\hfil}} \\

\cline{1-7}
\vbox to1.63ex{\vspace{1pt}\vfil\hbox to10ex{\hfil $S_1$\hfil}} & 
\vbox to1.63ex{\vspace{1pt}\vfil\hbox to8ex{\hfil 1\hfil}} & 
\vbox to1.63ex{\vspace{1pt}\vfil\hbox to8ex{\hfil 1\hfil}} & 
\vbox to1.63ex{\vspace{1pt}\vfil\hbox to8ex{\hfil 1\hfil}} & 
\vbox to1.63ex{\vspace{1pt}\vfil\hbox to8ex{\hfil 1\hfil}} & 
\vbox to1.63ex{\vspace{1pt}\vfil\hbox to8ex{\hfil 2\hfil}} & 
\vbox to1.63ex{\vspace{1pt}\vfil\hbox to8ex{\hfil -\hfil}} \\


\vbox to1.63ex{\vspace{1pt}\vfil\hbox to10ex{\hfil $S_2$\hfil}} & 
\vbox to1.63ex{\vspace{1pt}\vfil\hbox to8ex{\hfil 1\hfil}} & 
\vbox to1.63ex{\vspace{1pt}\vfil\hbox to8ex{\hfil 1\hfil}} & 
\vbox to1.63ex{\vspace{1pt}\vfil\hbox to8ex{\hfil 1\hfil}} & 
\vbox to1.63ex{\vspace{1pt}\vfil\hbox to8ex{\hfil 1\hfil}} & 
\vbox to1.63ex{\vspace{1pt}\vfil\hbox to8ex{\hfil 2\hfil}} & 
\vbox to1.63ex{\vspace{1pt}\vfil\hbox to8ex{\hfil -\hfil}} \\
\cline{1-7}
\end{tabular}
\caption{The field content under an $SU(2)_L\times SU(2)_H$ gauge and $SU(2)_R\times SU(2)_g\times U(1)_R$ global symmetries. There is also an accidental $Z_2$ symmetry 
with fields transforming as shown. The $U(1)_Y$ subgroup of $SU(2)_R$ is gauged. \label{tab1}}
\end{table}

The tree-level superpotential is given as\footnote{The terms with coefficients $y_3$ and $y_4$ were not included in Ref.~\cite{Harnik:2003rs}. These terms are not forbidden by any 
symmetries so we include them for completeness.} $W_{\rm FHtot} = W_1+W_2+W_3$ where
\bea
W_1&=&y_1S_1T_1T_2+y_2S_2T_3T_4+y_3S_1T_3T_4+y_4S_2T_1T_2\\
W_2&=&-mT_5T_6\\
W_3&=&y_5\left(\begin{array}{cc} T_1 & T_2 \end{array} \right)P \left( \begin{array}{c} T_5\\ T_6 \end{array} \right)+
y_6\left ( \begin{array}{cc} T_3 & T_4 \end{array} \right)Q \left( \begin{array}{c} T_5\\ T_6 \end{array} \right).
\label{treesup}
\eea

The $P$ and $Q$ mixing terms are there to marry off unwanted ``spectator" states such that the low energy effective theory is as minimal as possible. It is also possible to apply a $Z_3$ which protects us from tadpole terms involving either of the singlet fields, $S_1$ and $S_2$~\cite{Harnik:2003rs}. This $Z_3$ will commute with the existing symmetries. 

The gauge symmetry $SU(2)_H$ becomes strongly coupled at some scale, $\La_H$. Below $\La_H$, the appropriate degrees of freedom are mesons which are composite 
objects consisting of two ``T" doublets in the form $M_{ij}=T_iT_j$, with (i, j=1...6). There is a dynamically generated superpotential of the form Pf$M/\La_H^3$ as well as the tree 
level superpotential which follows from \eq{treesup}. As $P,Q, S_1$ and $S_2$ are not charged under $SU(2)_H$, they remain fundamental below $\La_H$. The canonically normalised effective 
superpotential reads
\bea
\nonumber
W_{dyn} &=&\la\left(\mbox{Pf}M-v_0^2M_{56}\right)+m_1S_1M_{12}+m_2S_2M_{34}+m_3S_1M_{34}+m_4S_2M_{12}\\
\nonumber
 &+& m_5\left(M_{15}P_{11}+M_{16}P_{12}+M_{25}P_{21}+M_{26}P_{22}\right)\\&+&m_6\left(M_{35}Q_{11}+M_{36}Q_{12}+M_{45}Q_{21}+M_{36}Q_{22}\right),
\eea
where, using Naive Dimensional Analysis (NDA) \cite{luty}, we have
\bea
v_0^2&\sim&\frac{m\La_H}{(4\pi)^2},\\
m_i&\sim& y_i\frac{\La_H}{4\pi},\\
\la(\La_H)&\sim&4\pi.
\eea

We now make the assumption that $(m_5, m_6)\gg (m_1, m_2, m_3, m_4)$, by a factor of 10 or so, and integrate out everything with a mass proportional to $m_5$ or 
$m_6$. This leaves us with a superpotential of the form
\bea
\nonumber
W^{\prime}_{dyn} &=&\la M_{56}\left(M_{14}M_{23}-M_{24}M_{13}-v_0^2+M_{12}M_{34}\right)\\
&+& m_1S_1M_{12}+m_2S_2M_{34}+m_3S_1M_{34}+m_4S_2M_{12}.
\eea

Assuming that $m_1\sim m_2 \sim m_3 \sim m_4 \sim m^{\prime}$, the fermionic components of the superfields, $S_1, S_2, M_{12}$ and $M_{34}$, mix and, provided 
$m_1m_2\ne m_3m_4$, the lightest eigenvalue of this mass matrix will generically have a mass of order $m^{\prime}$.

If we now do this diagonalization and integrate out all but the lightest eigenvalue, call it $S$, of the $S_1, S_2, M_{12}, M_{34}$ mass matrix, we are left with the 
superpotential
\beq
W=\la N\left(H_uH_d-v_0^2\right)+\frac{\la^{\prime}}{2}NS^2+\frac{m_{\tilde{s}}}{2}S^2,
\eeq
where we have changed notation according to the identifications
\beq
\left(\begin{array}{c} H_u^+ \\ H_u^0 \end{array} \right)=\left(\begin{array}{c} M_{13} \\ M_{23} \end{array} \right), \left(\begin{array}{c} H_d^0 \\ H_d^- \end{array} 
\right)=\left(\begin{array}{c} M_{14} \\ M_{24} \end{array} \right), N=M_{56}.
\label{lowesup}
\eeq
The parameter $\la^{\prime}=\la U_{ij}U_{kl}$, where $U_{ij}U_{kl}$ are components of the unitary matrix that diagonalizes the $S_1, S_2, M_{12}, M_{34}$ fermion mass 
matrix. The indices on the $U$s are there for show, the basis is irrelevant as we do not really care about the specific mixing between states.

The final assumption we make is that $m_{\tilde{s}}$ is parametrically larger than the electroweak scale and soft supersymmetry breaking masses. The superpotential in \eq{lowesup} is of 
the form we need with an additional linear term for the superfield $N$. This term is harmless with respect to the dark matter dynamics but we include it for completeness.

Ignoring the $S$ field for now, the remaining superpotential is that of the Fat Higgs model and the analysis of the electroweak vacuum structure proceeds as outlined in Ref.~\cite{Harnik:2003rs}. It is worth comparing the superpotential in \eq{lowesup} with that of the MNSSM \cite{pil1, Ded1}. In particular, the $N$ linear term in \eq{lowesup} is analogous to the tadpole terms appearing in the superpotential of Eq.(3.1) of Ref.~\cite{pil1}. In fact, the superpotential in \eq{lowesup} (apart from the $S$ terms) is that of the MNSSM. Consequently we can use the rather more detailed analysis of Refs.~\cite{pil1, Ded1} for the Higgs sector. 

The $S$ terms in \eq{lowesup} do not spoil the electroweak structure of the MNSSM. We can see this by integrating out $S$ using the equations of motion
\beq
\frac{\partial W}{\partial S}=\la^{\prime}NS+m_{\tilde{s}}S=0, \hspace{5mm} \Rightarrow  \hspace{5mm} S=0.
\eeq
Substituting the solution back into \eq{lowesup} we have the effective superpotential
\beq
W_{\rm eff}=\la N\left(H_uH_d-v_0^2\right),
\eeq
which is exactly the same as the superpotential for the MNSSM and the Fat Higgs model.


\section{Conclusions}
\label{conclusions}

In this article, we have discussed models in which a very heavy (3-30 TeV) dark matter candidate is present. In particular, we have focused on models motivated by Higgs Portal and Hidden Valley models, in which the dark matter (and the rest of the partially hidden sector) interacts with the Standard Model and its superpartners only through Higgs interactions.  

Dark matter annihilations in this scenario are considerably enhanced by non-perturbative contributions known as the Sommerfeld effect. Through this enhancement, dark matter particles with masses well above the electroweak scale can be produced thermally in the early universe with an abundance consistent with the measured density of dark matter.
The dark matter particle in this scenario, although well beyond the reach of the Large Hadron Collider, is still potentially detectable by direct and indirect dark matter experiments. Although we leave the details of this to future work~\cite{companion}, we point out that Sommerfeld corrections can dramatically enhance the dark matter annihilation rate in low velocity dispersion environments, such as dwarf spheriodal galaxies, thus considerably improving the prospects for indirect dark matter searches.

The particular model we study, which adds two extra SM singlet states to the MSSM spectrum, is independently motivated by the desire to raise the upper bound on the lightest higgs mass, thus lessening the LEP fine-tuning constraints.  We also showed that our model may be given a UV completion in the form of the previously considered  ``Fat Higgs'' models, where the singlet states are composites arising from strong hidden-sector dynamics.

Thus, Higgs Portal Dark Matter provides an example of an attractive and motivated alternative to conventional MSSM neutralino dark matter which is less fine-tuned and may be tested by current and future indirect detection experiments.

\bigskip
\bigskip

{\bf Acknowledgements} JMR and SMW are partially supported by the EC Network 6th Framework Programme Research and Training Network ``Quest for Unification" (MRTN-CT-2004-503369) and by the EU FP6 Marie Curie Research and Training Network ``UniverseNet" (MPRN-CT-2006-035863). DC is supported by the Science and Technology Facilities Council. DH is supported by the United States Department of Energy and NASA grant NAG5-10842. Fermilab is operated by the Fermi Research Alliance, LLC under Contract No. DE-AC02-07CH11359 with the United States Department of Energy. We would like to thank Markus Ahlers, Joe Silk, and Tim Tait for useful discussions. 


\end{document}